\documentclass[11pt]{article}
\newfont{\bbbold}{msbm10 scaled \magstep1}

\def\cL{{\cal L}}

\def\cN{{\cal N}}
\def\cO{{\cal O}}

\newfont{\goth}{eufm10 scaled \magstep1}

\def\a{\alpha}
\def\b{\beta}
\def\c{\gamma}\def\C{\Gamma}
\def\d{\delta}
\def\e{\epsilon}\def\ve{\varepsilon}
\def\f{\phi}
\def\h{\eta}

\def\l{\lambda}\def\L{\Lambda}

\def\th{\theta}

\def\be{\begin{equation}}\def\ee{\end{equation}}
\def\bea{\begin{eqnarray}}\def\eea{\end{eqnarray}}
\def\barr{\begin{array}}\def\earr{\end{array}}

\def\o{\omega}\def\O{\Omega}
\def\del{\partial}

\def\xz{\times}

\let\la=\label

\def\nn{\nonumber}
\def\bd{\begin{document}}
\def\ed{\end{document}}
\def\ba{\begin{array}}
\def\ea{\end{array}}
\def\bea{\begin{eqnarray}}
\def\eea{\end{eqnarray}}
\def\ft#1#2{\tfrac{#1}{#2}}
\def\fft#1#2{\frac{#1}{#2}}
\def\sst#1{{\scriptscriptstyle #1}}
\def\oneone{\rlap 1\mkern4mu{\rm l}}

\newcommand{\eq}[1]{(\ref{#1})}
\newcommand{\w}[1]{\\[0.#1cm]}
\def\eqs#1#2{(\ref{#1}-\ref{#2})}
\def\det{{\rm det\,}}
\def\tr{{\rm tr}}

\newcommand{\hoch}[1]{$\, ^{#1}$}
\newcommand{\imperial}{\it\small Theoretical Physics Group, Imperial College London\\ Prince Consort Road, London SW7 2AZ, UK}
\newcommand{\kings}
{\it\small Department of Mathematics, King's College, University of London\\ Strand, London WC2R 2LS, UK}
\newcommand{\uu}
{\it\small Department of Theoretical Physics, Uppsala, Sweden}
\newcommand{\hip}
{\it\small HIP-Helsinki Institute of Physics, P.O. Box 64 FIN-00014
University of Helsinki, Suomi-Finland}
\newcommand{\stock}
{\it\small Department of Theoretical Physics, Stockholm, Sweden}
\newcommand{\golm}
{\it\small AEI, Max Planck Institut f\"ur Gravitationsphysik\\ Am M\"{u}hlenberg 1, D-14476 Potsdam, Germany}
\makeatletter
\@addtoreset{equation}{section} \makeatother

\newcommand{\sa}{/ \hspace{-1.2ex}}
\newcommand{\saa}{/ \hspace{-1.4ex}}
\newcommand{\saaa}{\, / \hspace{-1.6ex}}
\newcommand{\Scal}[1]{\Bigl ({#1} \Bigr )}
\newcommand{\scal}[1]{\bigl ({#1} \bigr )}

\newcommand{\CR}{\nonumber \\*}

\newcommand{\trace}{\hbox {tr}~}
\newcommand{\traceS}{\hbox {tr}_{\scriptscriptstyle \mathfrak{S}}~}

\DeclareMathAlphabet{\mathpzc}{OT1}{pzc}{m}{it}
\def\BRST{\,\mathpzc{s}\,}
\def\aBRST{{\scriptstyle (\mathpzc{s})}}
\def\q{{{\scriptscriptstyle (Q)}}}
\def\qs{{\scriptscriptstyle (Q\mathpzc{s})}}
\def\Qsla{{\mathcal{S}_{\q}}}
\def\Slav{{\mathcal{S}_\aBRST}}
\def\epsilonb{{\overline{\epsilon}}}
\def\bulletup{{\scriptstyle \bullet}}

\newcommand{\gra}[2]{{\scriptscriptstyle (#1 , #2 )}}
\newcommand{\ord}[1]{{\scriptscriptstyle (#1)}}

\def\cL{{\cal L}}
\def\cN{\mathcal{N}}
\def\cO{\mathcal{O}}

\def\ie{{\it i.e.}\ }
\def\eg{{\it e.g.}\ }

\newcommand{\sfrac}[2]{{\scriptstyle \frac{#1}{#2}}}
\newcommand{\stfrac}[2]{{\scriptscriptstyle \frac{#1}{#2}}}

 \def\balpha{{\overline{\alpha}}}
 \def\bbeta{{\overline{\beta}}}
 \def\bgamma{{\overline{\gamma}}}
 \def\bdelta{{\overline{\delta}}}
 \def\bepsilon{{\overline{\epsilon}}}
 \def\bvarepsilon{{\overline{\varepsilon}}}
 \def\bzeta{{\overline{\zeta}}}
 \def\bareta{{\overline{\eta}}}
 \def\btheta{{\overline{\theta}}}
 \def\bvartheta{{\overline{\vartheta}}}
 \def\biota{{\overline{\iota}}}
 \def\bkappa{{\overline{\kappa}}}
 \def\blambda{{\overline{\lambda}}}
 \def\bmu{{\overline{\mu}}}
 \def\bnu{{\overline{\nu}}}
 \def\bxi{{\overline{\xi}}}
 \def\bpi{{\overline{\pi}}}
 \def\brho{{\overline{\rho}}}
 \def\bvarrho{{\overline{\varrho}}}
 \def\bsigma{{\overline{\sigma}}}
 \def\bvarsigma{{\overline{\varsigma}}}
 \def\btau{{\overline{\tau}}}
 \def\bphi{{\overline{\phi}}}
 \def\bvarphi{{\overline{\varphi}}}
 \def\bchi{{\overline{\chi}}}
 \def\bpsi{{\overline{\psi}}}
 \def\bomega{{\overline{\omega}}}

\def\thalf{{\textrm{\tiny\textonehalf}}}
\def\tquarter{{\textrm{\tiny\textonequarter}}}
\def\Ko{{\scriptscriptstyle K}}
\def\tKo{\scriptscriptstyle k }
\def\corr{$\clubsuit$}

\newcommand{\auth}{\large G.\ Bossard\footnote{email: bossard@aei.mpg.de}, P.S.\ Howe\footnote{email: paul.howe@kcl.ac.uk} and K.S.\ Stelle\footnote{email: k.stelle@imperial.ac.uk}}

\begin{document}

\renewcommand{\thefootnote}{\fnsymbol{footnote}}

\null
\begin{flushright}
{\small AEI-2009-087}\\
{\small KCL-MTH-09-09}\\
{\small Imperial/TP/09/KSS/03}
\vskip 1.5 cm
\end{flushright}

\begin{center}
{\Large{\bf A note on the  UV behaviour of maximally supersymmetric Yang-Mills theories}}
\vspace{.75cm}

\auth

\vspace{.5cm}

$^{\ast}$\golm
\\
\vskip 1 em
$^{\dagger}$\kings
\\
\vskip 1 em
$^{\ddagger}$\imperial

\vspace{1cm}

{\bf Abstract}
\end{center}
\vskip .5cm
The question of whether BPS invariants are protected in maximally supersymmetric Yang-Mills theories is investigated from the point of view of algebraic renormalisation theory. The protected invariants are those whose cohomology type differs from that of the action. It is confirmed that one-half BPS  invariants ($F^4$) are indeed protected while the double-trace one-quarter BPS invariant ($d^2F^4$) is not protected at two loops in $D=7$, but is protected at three loops in $D=6$ in agreement with recent calculations. Non-BPS invariants, i.e. full superspace integrals, are also shown to be unprotected.

\vspace{1cm}

\null

\renewcommand{\thefootnote}{\arabic{footnote}}
\setcounter{footnote}{0}

\pagebreak
\setcounter{page}{1}

An intriguing question in maximally supersymmetric theories is which F-terms (or BPS invariants) are protected from UV divergences and which are not. Superspace non-renormalisation theorems in conventional superspace \cite{Howe:1988qz} allow the possibility of one-half BPS counterterms, i.e. integrals over eight odd coordinates ($\th$s) for maximally supersymmetric Yang-Mills theories (MSYM), a prediction which was in agreement with the old Feynman diagram computations of ref \cite{Marcus:1984ei}. The more efficient unitarity methods \cite{Bern:1994zx} have allowed computations to be carried out at much higher loop order, however, and in 1998 there were indications that MSYM could be finite at $L=4$ loops in $D=5$ \cite{Bern:1998ug}, despite the existence of an eight-$\th$ invariant. This expectation has now been confirmed \cite{Bern:2006ew,BernYM} and shows that conventional superspace methods are not sufficiently powerful to account fully for the UV behaviour.

It seemed that this problem could be circumvented by means of off-shell harmonic superspace methods. There is an off-shell version of $N=3, D=4$ SYM \cite{Galperin:1985uw}
(which has the same physical spectrum as $N=4$) which one would naively expect to forbid one-half BPS counterterms but admit one-quarter BPS ones \cite{Howe:2002ui}, i.e. integrals over twelve $\th$s. This would then explain the $D=5, L=4$ MSYM result and is also compatible with the one-quarter BPS divergence found at $L=2$ in $D=7$ \cite{Marcus:1984ei}. There is also an off-shell version of MSYM with a finite number of auxiliary fields which preserves nine supersymmetries (one-half-susy-plus-one) which would seem to lead to the same predictions \cite{Baulieu:2007ew,Bossard:2009sy}. This year, however, unitarity computations have revealed that the double-trace one-quarter BPS invariant, although divergent in $D=7,L=2$, is actually finite in $D=6,L=3$ \cite{BernYM}, a result which is at odds with the expectation that only one-half BPS invariants are protected \cite{Bossard:2009sy}.

A possible explanation for the failure of extended superfield methods to account for this result, along with a similar one for $D=5$ maximal supergravity (MSG) at $L=4$ \cite{Bern:2009kd}, is that both of the off-shell formulations referred to above do not preserve all of the other symmetries. Both break manifest Lorentz invariance and R-symmetry while the remaining supersymmetries are non-linearly realised. It is possible that the superspace non-renormalisation theorems could be improved by taking this feature into account, but it is a difficult problem. On the other hand, the algebraic approach advocated in \cite{Bossard:2009sy} has the advantage that all of the symmetries are kept under tight control even though it eschews the use of auxiliary fields. In this note we shall show that a closer examination of the algebraic version of the supersymmetry non-renormalisation theorem leads to the result that the double-trace one-quarter BPS invariant is indeed protected in $D=6$ even though it is not in $D=7$. The implication of this is that the result of reference \cite{BernYM} is explicable in terms of the obvious symmetries of MSYM, although this has yet to be extended to the finiteness of $D=4,L=5$ MSG. The field theory predictions for the onset of UV divergences for MSYM are thus in agreement with existing calculations as well as with the recent predictions made from a string theory viewpoint \cite{Berkovits:2009aw}.

Regarding the use of string theory to make predictions about the UV behaviour of MSYM or MSG field theories, we would like to recall the known difficulties in using systems with infinite numbers of extra fields as field-theory ``regulators.'' This was clearly pointed out in the Kaluza-Klein context in reference \cite{Duff:1982gj}, where, using zeta-function regularisation, it was shown how, despite the decoupling of individual KK massive modes in a compactification limit, there can nonetheless be divergence cancellations that take place between the lower-dimensional theory to be ``regulated'' and the contributions arising from the infinity of KK massive modes. For example, odd-loop-order gravity or supergravity divergences in odd numbers of spacetime dimensions vanish owing to the absence of available Lorentz and diffeomorphism invariant counterterms. But this does not imply that the massless KK sectors in even, lower dimensionalities are free of divergences, merely that such divergences cancel against the summed effects of the ``regulators.'' It would be nice to understand how this problem is circumvented by the use of string theory with its doubly infinite numbers of ``regulators'' in the string and KK massive states. This issue would appear to call into question the usefulness of string theory as a quantum regulator for a lower-dimensional field theory that occurs as its zero slope limit, unless there is some reason why classical truncation consistency is preserved at the quantum level. It might be, perhaps, that the case of maximally supersymmetric theories is special in this context.

Before starting on the details it is worth recalling what the leading bosonic contributions to the four-point BPS invariants are for MSYM in spacetime. There are two one-half BPS invariants, $Tr(F^4)$ and $(Tr(F^2))^2$, and two one-quarter BPS ones, the single- and double-traces of $F^4$ with two extra spacetime derivatives. Both of the one-half BPS invariants are true BPS states in that they cannot be written as integrals of gauge-invariant integrands over more than eight $\th$s but the single-trace one-quarter BPS invariant is not. In fact, it can be written as the full superspace integral of the Konishi operator \cite{Drummond:2003ex} and is therefore non-protected in agreement with the computational results \cite{BernYM}.

The algebraic approach to the renormalisation of maximally supersymmetric theories was discussed in some detail in \cite{Bossard:2009sy}. Here we give a brief synopsis of the method. The basic idea is to study the symmetry properties of the effective action $\C$ algebraically. In the absence of any convenient set of auxiliary fields it is best to discard them completely and to work in components. The supersymmetry transformations are then non-linear, the algebra only closes modulo gauge transformations and the equations of motion, and gauge-fixing is not manifestly supersymmetric. All of these technical problems can be overcome by the  Batalin-Vilkovisky (BV) version of standard BRST techniques \cite{Dixon:1990jv,Howe:1990pz,Baulieu:2006gx}. An important point is that one needs to introduce a ``supersymmetry ghost'', which is a constant commuting spinor $\epsilon$ in the case of rigid supersymmetry.\footnote{This becomes the Faddeev-Popov ghost of local supersymmetry in supergravity.} One can then show that, in addition to the BRST operator $s$  (of ghost number one) associated to gauge invariance, there is an additional operator $Q$ (with which we associate one unit of a new type of ghost number called shadow number) under which any putative counterterm should be invariant too. $Q$ acts as a supersymmetry transformation with parameter $\epsilon$ on gauge-invariant functions of the fields and their derivatives in the cohomology of $s$, and satisfies\footnote{The $\approx$ symbol refers to the fact that identity holds modulo the equations of motion in the physical sector.}

\be
Q^2 \approx  -\pounds_v\ ,  \label{nilpo} \ee

where $v^a:=-\frac{i}{2}\bar\e\C^a\e$.
If we express an invariant as an integral of a spacetime $D$-form, $\cL_D$ say, then we have

\be
 Q \cL_D + d_0 \cL_{D-1,1} \approx 0
 \la{1}
\ee

where $\cL_{D-1,1}$ is a spacetime $(D-1)$-form linear in $\epsilon$ (and thus with shadow number one) and $d_0$ is the spacetime exterior derivative. Applying $Q$ to \eq{1} and using (\ref{nilpo}) and the fact that it anticommutes with $d_0$ we deduce that

\be
 Q \cL_{D-1,1} + d_0 \cL_{D-2,2} + i_v \cL_D  \approx 0
 \la{2}
\ee

and so on (where $ i_v$ is the contraction operator, $ i_v dx^a = v^a$). Thus we obtain a cocycle of the extended differential $\tilde d :=d_0 + Q +i_v $ whose components $\cL_{D-q,q}$ are $(D-q)$-forms with shadow number $q$. Now the question of whether a given invariant is required as a counterterm, i.e. corresponds to a UV divergence, can be reformulated in terms of the anomalous dimension of the same invariant considered as a composite operator insertion, by use of the Callan-Symanzik equation \cite{Sor,Bossard:2009sy}. Furthermore, we can include all of the terms in the cocycle as operator insertions for any invariant including the original starting action. We can therefore conclude that an invariant will be a required counterterm if it has the same cocycle structure as the initial action. This is the the generalisation of the  algebraic supersymmetry non-renormalisation theorem \cite{Sor,Sor4,GuillaumeN4}, to non-renormalisable theories \cite{Bossard:2009sy}.

By a slight extension of the theorem of \cite{Henneaux}, the cohomology of the BRST operator $s$ of ghost number zero and shadow number $q$ corresponds to the gauge invariant functions of the fields of order $q$ in the constant spinor $\epsilon$, identified modulo the equations of motion.
So a term $\cL_{D-q,q}$ in a cocycle corresponds to a $(D-q)$-form with $q$ additional spinor indices which have to be totally symmetrised as $\epsilon$ is a commuting object. This implies that the cocycle is equivalent to a closed $D$-form in superspace. We can therefore study the possible solutions to the algebraic non-renormalisation problem systematically using superspace cohomology. Indeed, from a computational point of view, one has the following identifications between objects in superspace and in components

\be   d_1 \sim Q \hspace{10mm}Êt_0 \sim i_v  \hspace{10mm} d \theta^\alpha \sim \epsilon^\alpha \hspace{10mm}d\theta^\alpha ÊA_\alpha  \sim c  \label{equival} \ee

where $c$ is the shadow field \cite{Baulieu:2006gx} and the superspace objects are defined below. This is advantageous because it allows us to study the problem starting at the lowest dimension and work upwards rather than the other way round. Since the top component has many terms besides the leading bosonic one this can be a rather complicated object to construct. Of course, any invariant can also be presented as a superspace integral, and in general the superfield integrand will have many more components than appear in the cocycle, so it seems that the algebraic approach implies that the essential part of a superfield integrand is actually the part that appears in the closed super $D$-form. For example, as we shall see later, the cocycle associated with a one-half BPS invariant is actually longer than the cocycle for the action, whereas the cocycle associated with a full superspace integral is the same as that for the action.

It has been known for some time that one can write a supersymmetric invariant as a spacetime integral in terms of a closed super-form, a procedure which has been dubbed ``ectoplasm''\cite{Gates:1997kr,Gates:1997ag}. Suppose $M$ is a supermanifold with $D$-dimensional body $M_0$ and $L_D$ is a closed $D$-form on $M$. The formula for an invariant $I$ is

\be
 I=\int_{M_0}\, L_{D,0}(x,\th=0)\ ,
 \la{3}
\ee

where $L_{D,0}$ is the purely bosonic component of $L_D$ with respect to some coordinate basis and where $(x,\th)$ are (even, odd) coordinates on $M$. It is easy to see that this does give a supersymmetry invariant because, under an infinitesimal diffeomorphism of $M$, a closed form changes by a total derivative, and a spacetime supersymmetry transformation is given by the leading term of an odd superdiffeomorphism in its $\th$-expansion. Since an exact $D$-form integrates to zero, it follows that we need to analyse the $D$th  de Rham cohomology group of $M$ in order to find the possible invariants. This has nothing to do with topology, however, since the forms we are interested in have components which are gauge-invariant functions of the physical fields and this leads to non-trivial cohomology even for flat supermanifolds.

We now give a brief review of some essential aspects of superpace cohomology. We shall only consider flat superspace here. The standard superinvariant basis one-forms are

\bea
 E^a&=& d x^a-\frac{i}{2} d\th^\a (\C^a)_{\a\b} \th^\b \nn\w1
 E^\a&=& d\th^\a \ ,
 \la{4}
\eea

which are dual to the usual invariant derivatives $(\del_a, D_\a)$. As we are going to focus on MSYM the index $\a$ can be thought of as a $16$-component $D=10$ chiral spinor index, although in $D<10$ it will stand for a combined spinor and R-symmetry index. Similarly, $\C^a$ denotes the ten-dimensional gamma matrices which reduce to a direct product of internal and spinor matrices.

The fact that the tangent spaces of a superspace (even in the curved case) split invariantly into even and odd subspaces implies that one can introduce a bi-grading on the spaces $\O^n$ of differential $n$-forms, $\O^n=\oplus_{p+q=n} \O^{p,q}$. We can also split the exterior derivative $d$ into the following components with the indicated bidegrees \cite{Bonora:1986ix}

\be
 d=d_0(1,0) + d_1(0,1) + t_0(-1,2)\ .
 \la{5}
\ee

In a general superspace there is also a component $t_1$ of bidegree $(2,-1)$ but it vanishes in flat space (and does not play a crucial cohomological role in any case). $d_0=E^a \del_a$ and $d_1=E^\a D_\a$ are respectively even and odd exterior derivatives, while $t_0$ is an algebraic operation involving the dimension zero torsion, which is proportional to $\C$. For $\o\in\O^{p,q}$,

\be
 (t_0\o)_{a_2\ldots a_p\b_1\ldots \b_{q+2}}\sim (\C^{a_1})_{(\b_1\b_2} \o_{a_1\ldots a_p\b_{3}\ldots \b_{q+2})}\ .
 \la{6}
\ee

Since $d^2=0$ we find, amongst other relations,

\bea
 t_0^2&=&0 \la{7}\w1
 t_0 d_1 + d_1 t_0&=& 0 \la{8}\w1
 d_1^2 + t_0 d_0 + d_0 t_0&=&0 \ . \la{9}
\eea

Equation \eq{7} implies that we can define $t_0$-cohomology groups $H_t^{p,q}$ \cite{Bonora:1986ix}. We can then define a new odd derivative $d_s$ acting on elements of these groups by

\be
 d_s [\o]:= [d_1 \o]\ ,
 \la{10}
\ee

where $\o\in[\o]\in H_t^{p,q}$, with $[\o]$ denoting the cohomology class of a $t_0$-closed form $\o$. Equations \eq{8} and \eq{9} then imply that these definitions are independent of the choice of representative $\o$ and that $d_s^2=0$. This means that we can define the so-called spinorial cohomology groups $H_s^{p,q}$\cite{Cederwall:2001dx,Howe:2003cy}. The point of these definitions is that they enable us to solve for the superspace cohomology of $d$ in terms of the spinorial cohomology groups. Specifically, suppose the lowest-dimensional non-zero component (i.e. the one with the largest number of odd indices) of some closed $D$-form $L_D$ is $L_{D-q,q}$, for some $q$, then, since $dL_D=0$, we have $t_0 L_{D-q,q}=0$, and since we are interested in cohomology, the starting component will correspond to an element of $H_t^{D-q,q}$. The next component of $dL_D=0$ then tells us that $d_s [L_{D-q,q}]=0$. Thereafter, if we can solve this equation, we can solve for all of the higher components of $L_D$ in terms of $L_{D-q,q}$.\footnote{In principle there can be higher-order obstructions but these do not arise in the examples discussed here.} There may, of course, be other solutions to the problem with lowest components of different bidegrees, but this is precisely what is needed for there to be non-trivial examples of non-renormalisation theorems as this implies the existence of more than one type of cocycle. Another important consideration is that any putative lowest component of a closed $D$-form must lead to a non-zero $L_{D,0}$.

We shall now discuss the cohomology of $N=1, D=10$ superspace (see \cite{Berkovits:2008qw}
for more details). Interestingly enough, it turns out to be closely related to the pure spinor approach to supersymmetry \cite{Howe:1991mf,Howe:1991bx}. Consider first $H_t^{0,q}$. Let $\o\in \O^{0,q}$ and let $\bar\o:= u^{\a_q}\ldots u^{\a_1} \o_{\a_1\ldots \a_q}$ where $u$ is a (commuting) pure spinor, $u\C^a u=0$. Clearly, if $\o\mapsto \o + t_0 \l$, where $\l\in \O^{1,q-2}$, $\bar\o$ is unchanged, so that $H_t^{0,q}$ is isomorphic to the space of $q$-fold pure spinors which appears in pure spinor cohomology \cite{Berkovits:2002zk}. The $t_0$ cohomology groups for $1\leq p\leq 5$ are again spaces of pure spinor type objects but with additional antisymmetrised vector indices. This arises because of the gamma-matrix identities which are responsible for the kappa-symmetry of the string and fivebrane actions. In form notation these are

\be
 t_0\C_{1,2}=t_0 \C_{5,2}=0
 \la{11}
\ee

where $\C_{p,2}$ denotes a symmetric gamma-matrix with $p$ even indices viewed as a $(p,2)$-form. For our problem only the second of these relations is relevant. For example, suppose $\o\in \O^{3,q}$ can be written

\be
 \o_{3,q}=\C_{5,2}\l^2{}_{,q-2}\ ,
 \la{12}
\ee

where the notation indicates that two of the even indices on $\C_{5,2}$ are to be contracted with the two vector indices on $\l$, then it is clearly the case that $\o$ is $t_0$-closed. Furthermore, in cohomology, the object $\l$ can be taken to be of pure spinor type on its odd indices. Constructions such as this are not possible for $p\geq 6$ and it turns out that all such $t_0$-cohomology groups vanish.

Although it would seem that there are quite a lot of cohomology groups available which one might consider as possible lowest components for closed $D$-forms it turns out that there is only one type of cocycle in $N=1, D=10$, with lowest component $L_{5,5}$ \cite{Berkovits:2008qw}. This is due to the fact that this is the only case which can lead to a non-zero $L_{10,0}$. So any closed $D$-form in $D=10$ superspace has a lowest component of the form

\be
 L_{5,5}=\C_{5,2} M_{0,3}
 \la{13}
\ee

where $d_s[M_{0,3}]=0$. The simplest example of this is for an unconstrained scalar superfield $S$, which corresponds to a full superspace integral,

\be
 M_{\a\b\c}= T_{\a\b\c,\d_1\ldots \d_5} D^{11 \d_1\ldots \d_5} S \ ,
 \la{14}
\ee

where $T$ is an invariant tensor constructed from gamma-matrices \cite{Berkovits:2002zk} and $D^{11\a_1\ldots \a_5}$ is the dual of the antisymmetrised product of eleven $D_\a$s. The tensor $T$ is symmetric on $\a\b\c$ and totally antisymmetric on the $\d$s. Now a closed $D$-form in $D$ dimensions gives rise to a closed $(D-1)$-form in $(D-1)$ dimensions under dimensional reduction, so this means that we can immediately construct the cocycle associated with any non-BPS invariant in $4<D<10$; it will have lowest component $L_{D-5,5}\sim\C_{D-5,2} M_{0,3}$ where $\C_{D-5,2}$ is the dimensional reduction of $\C_{5,2}$.

The next example we shall consider is the (on-shell) action. It is an example of a Chern-Simons (CS) invariant. In $D$ dimensions such an invariant can be constructed starting from a closed, gauge-invariant $(D+1)$-form $W_{D+1}=dZ_D$, where $Z_D$ is a potential $D$-form \cite{Howe:1998tsa}, provided that it has the property of Weil triviality \cite{Bonora:1986xd}, i.e. it can also be written as $dK_D$ for some gauge-invariant $D$-form $K_D$. If this is true, then $L_D:=K_D-Z_D$ is closed and can be used to construct an integral invariant via the ectoplasm formula. For the $D=10$ SYM action the appropriate $W_{11}$ is $H_7 Tr(F^2)$ where, in flat superspace, the closed seven-form $H_7\sim \C_{5,2}$. This eleven-form is easily seen to have the correct property, with the lowest component of $K_D$ being $K_{8,2}$; $Z$ can be chosen to be $H_7 Q_3$ where $Q_3$ is the SYM Chern-Simons three-form, $d Q_3=Tr(F^2)$. The lowest non-zero component of $L_{10}$ is

\be
 L_{5,5}=-\C_{5,2} Q_{0,3}\ .
 \la{15}
\ee

We can again reduce this formula to any dimension $4<D<10$, and conclude that the lowest term in the closed $D$-form associated with the action, or action-form, for all of these cases is $L_{D-5,5}=-\C_{D-5,2} Q_{0,3}$. This is in agreement with the cocycle obtained in components in \cite{Bossard:2009sy}, with the identifications (\ref{equival}). We are therefore able to conclude that the cocycle type of the action is the same as that of a non-BPS invariant in all dimensions $D\geq 5$, and therefore that such full superspace integrals are not protected by the algebraic non-renormalisation theorem.

We now move on to discuss the BPS invariants, starting with one-half BPS. There are two of these corresponding to single- and double-trace $F^4$ invariants. There is not a lot of difference between them from the point of view of superspace cohomology, and we shall focus on the double-trace as it will be useful in the subsequent discussion of the double-trace one-quarter BPS case. In $D=10$ this invariant is again of CS type with $W_{11}=H_3 F^4$, but the CS nature is lost for $D\leq 8$ due to the low rank of $H_3\sim \C_{1,2}$ and so we shall derive the associated closed $D$-form starting from scratch in $D=7$ and below.\footnote{$F^4$ arises at one loop in D=8 where it is divergent; this is compatible with both algebraic and superspace non-renormalisation theorems because they are not valid at one loop.}

In $D\leq 8$, the SYM field strength multiplet is a scalar superfield $W_r$, $r=1,\ldots n=10-D$, whose independent components are the physical scalars and spinors and the spacetime field strength, and from which one can construct two bilinear multiplets, the Konishi mulitplet $K:=Tr(W_r W_r)$, and the supercurrent, $J_{rs}:=Tr(W_r W_s)- \frac{1}{n}\d_{rs} K$. The supercurrent is itself one-half BPS, but it is ultra-short in the sense that its $\th$-expansion only goes up to $\th^4$ (as opposed to $\th^8$ for a standard one-half BPS superfield). It has 128+128 components while Konishi is an unconstrained scalar superfield in the interacting theory. The supercurrent contains all of the conserved currents of SYM: the R-currents, the supersymmetry current, the energy-momentum tensor and an identically conserved topological current for the gauge fields.

If we square $J$ we obtain scalar superfields in various representations of the R-symmetry group. The totally symmetric, traceless representation is the one-half BPS multiplet we are interested in. Let us consider first $D=7$. We can take the R-symmetry group to be $SU(2)$ and use $i,j$ etc to denote doublet $SU(2)$ indices. The supercurrent is $J_{ijkl}$, while the one-half BPS multiplet is $B_{i_1\ldots i_8}:=J_{(i_1\ldots i_4} J_{i_5\ldots i_8)}$. It obeys the constraint

\be
 D_{\a i} B_{j_1\ldots j_8}=\ve_{i(j_1} \L_{\a j_2\ldots j_8)}\ ,
 \la{16}
\ee

where the spinor index can take on 8 values. The lowest component of the associated closed seven-form is an $L_{0,7}$ of the form

\be
 L_{\a_1 i_1,\ldots,\a_7 i_7}=\h_{(\a_1\a_2}\ldots \h_{\a_5\a_6} \L_{\a_7) i_1\ldots i_7}\ ,
 \la{17}
\ee

where $\h_{\a\b}$ is the (symmetric) charge-conjugation matrix. It is straightforward to verify that this defines an element of $H_s^{0,7}$ and that it contains a singlet $L_{7,0}$, the spacetime double-trace $F^4$ invariant. Furthermore, it is not difficult to show that this seven-form cannot be brought to the same form as that of the action by the addition of some exact term. This shows that the one-half BPS invariant has a different cocycle structure to the action, although this fact is not directly relevant in $D=7$ as this counterterm cannot arise there anyway for dimensional reasons.

Now let us consider the one-quarter BPS double-trace invariant $d^2 F^4$. It turns out that it can be written as a subsuperspace integral of an associated pseudo-one-half BPS superfield which is constructed from the one above by the insertion of two contracted spacetime derivatives, one on each factor of $J$. This allows us to write down a candidate closed seven-form immediately with lowest component given as in \eq{17} but where now $\L\sim \del \chi \cdot\del J$ where $DJ\sim \chi$. In this case, however, one can show that

\be
 L_{0,7}=d_1 K_{0,6} + t_0 K_{1,5}
 \la{18}
\ee

for some $K_{0,6}$ and $K_{1,5}$ which are constructed explicitly in terms of bilinears in the components of $J$. This is enough to show that this closed seven-form is cohomologically equivalent to the action form as there are only two types of cocycle in $D=7$. Hence the one-quarter double-trace BPS invariant in $D=7$ is not protected.

The above closed seven-form can be reduced straightforwardly to give a closed six-form in $D=6$ which must also have the same cocycle structure as the action. One might therefore conclude that this invariant cannot be protected in $D=6$ either. However, the R-symmetry group in $D=7$ is $SU(2)$ while for $N=2, D=6$ it is $SU(2)\xz SU(2)$ and there is no guarantee that the reduced six-form will have the full R-symmetry. For this reason we shall analyse $N=2, D=6$ starting again from the supercurrent.

The $N=2, D=6$ supersymmetry algebra is

\bea
 \{D_{\a i}, D_{\b j}\}&=&i \ve_{ij} (\c^a)_{\a\b}\del_a\nn\w1
 \{D^{\a i'}, D^{\b j'}\}&=&i \ve^{i'j'} (\c^a)^{\a\b}\del_a\nn\w1
 \{D_{\a i}, D^{\b j'}\}&=& 0 \ ,
 \la{19}
\eea

where $\a=1\ldots 4$ is a chiral spinor index and $i,i'$ are doublet indices for the two $SU(2)$s. In this notation the field strength is $W_i{}^{i'}$ and the supercurrent is $J_{ij}^{i'j'}:= Tr( W_{(i}{}^{i'} W_{j)}{}^{j'})$. The double-trace true one half BPS superfield is $B_{ijkl}^{i'j'k'l'}:= J_{(ij}^{(i'j'} J_{kl)}^{k'l')}$. It obeys the constraint

\be
 D_{\a (i} B_{jklm)}^{j'k'l'm'}=0
 \la{20}
\ee

together with a similar one for the upper indices. The one-half BPS Lagrangian six-form starts at $L_{0,6}$. It is

\be
 L_{\a i\b j\c k}{}^{\d l'\e m'\f n'}:= \d_{(\a}{}^{(\d} \d_{\b}{}^\e B_{\c)ijk}^{\f)l'm'n'}\ ,
 \la{21}
\ee

where

\be
 B_{\a ijk}^{\b i'j'k'}:= D_\a{}^l D^\b_{l'} B_{ijkl}^{i'j'k'l'}\ .
 \la{22}
\ee

There are two $Spin(1,5)$ representations here, a singlet and a 15, but it turns out that precisely this combination is required in order to obtain an element of $H_s^{0,6}$. Moreover, it is not difficult to show that this form cannot be shortened so that the cocycle for the true one-half BPS invariant is different to that of the action (which starts at $L_{1,5}$ in $D=6$).

As in the $D=7$ case the double-trace one-quarter BPS invariant can be constructed in terms of a pseudo-one-half BPS superfield obeying \eq{20}. Again it is formed by inserting a pair of contracted spacetime derivatives, one on each factor of $J$. We now have the task of testing for the cohomological triviality of the corresponding closed six-form, i.e. we try to write $L_{0,6}=d_1 K_{0,5}+ t_0 K_{1,4}$. In contradistinction to the $D=7$ case, however, we find that we cannot do this.

The problem can be approached from different points of view. The first is to try repeat what was done for $D=7$ by writing $K$ in terms of bilinears of the supercurrent, but it turns out that no such $K_{0,5}$ and $K_{1,4}$ can be constructed in this way. Alternatively, we can observe that there are two true one-quarter BPS bilinears that can be constructed from $J$,

\bea
 C_{ijkl}&:=& J_{(ij}^{i'j'} J_{kl)i'j'}\nn\w1
 C'^{i'j'k'l'}&:=& J_{(ij}^{(i'j'} J^{ k'l')ij}\ .
 \la{23}
\eea

These superfields obey constraints of the type \eq{20}, $C$ with respect to $D$ and $C'$ with respect to $D'$. There is another shortened bilinear that can be constructed from $J$; it is

\bea
 S_{ij}^{i'j'}:= J_{k(i }^{k'(i'} J_{j) l}^{j')l'} \ve^{kl}\ve_{k'l'}\ .
 \la{24}
\eea

It obeys constraints that are third order in $D$ and $D'$ separately. It is akin to the product of two supercurrents in $N=4, D=4$ which is protected as a superconformal field even though it is not BPS-shortened \cite{Eden:2000bk,Heslop:2001dr}.

The pseudo-one-half BPS $B$ can be written as four derivatives on any of these three superfields, up to a total spacetime derivative which is irrelevant under integration. We have

\bea
 B_{ijkl}^{i'j'k'l'}&\sim& D^4_{ijkl} C'^{i'j'k'l'}\nn\w1
 &\sim& D^{4i'j'k'l'} C_{ijkl}\nn\w1
 &\sim& D^2_{\a\b (ij} D^{2\a\b (i'j'} S_{kl)}^{k'l')} \ ,
 \la{25}
\eea

where the $D^4$s are fourth-order in $D$ and totally symmetric on the internal indices, while the second-order $D^2$s are symmetric in the internal indices and antisymmetric on the spinor indices. The one-quarter BPS invariant can be written as a twelve-theta integral of any of these so that one might expect to trivialise the cohomology by using any one of them in $K$. But it turns out to be not possible even if one includes all three at once.

We therefore conclude that the double-trace $d^2 F^4$ invariant is protected in $N=2, D=6$ SYM even though a similar invariant is not protected in $D=7$. A key difference between the two cases is the larger R-symmetry group in $D=6$ which is more restrictive when it comes to constructing possible trivialising $(D-1)$-forms $K$.

This result is in agreement with the latest $D=6,L=3$ SYM calculations \cite{BernYM}. There is now only one remaining BPS counterterm to be checked in MSYM, the double-trace one-quarter BPS invariant which could appear in principle in $D=5$ at $L=6$ loops. Although we have not checked this explicitly, it seems likely that it will be protected because we can obtain a protected cocycle by dimensional reduction of the $N=2,D=6$ cocycle we have just discussed. This would not necessarily have the full $Sp(2)$ R-symmetry but it would have a larger $R$-symmetry than the trivial cocycle that can be constructed by dimensional reduction from $D=7$.

The evaluation of the various superspace cohomology groups for maximal supergravity is a more difficult problem, principally because of the larger R-symmetry groups, many of which have the disadvantage of being symplectic. There is also a conceptual issue to deal with because the precise relation between the cohomology problem in algebraic renormalisation in components and the ``ectoplasm'' cohomology problem in superspace has not yet been identified for supergravity. The equivalence between these two certainly does not hold for the cocycle associated to the classical action, since the latter vanishes on-shell. Nevertheless we can speculate as to the outcome of such investigations using MSYM as a guide. Let us suppose that cohomological arguments can be found which protect the $D=5, L=4$ invariant (which is one-eighth BPS); then, by dimensional reduction, we would expect this counterterm to be protected also in $D=4$ where it could occur at $L=6$ loops. There is still the question of the one-quarter BPS counterterm which could occur at $L=5$ in $D=4$, but it would seem unlikely that this would be divergent while the $L=6$ one is not. The net upshot of this is that it would seem likely that all BPS counterterms in $N=8,D=4$ supergravity are protected after all, and that the first divergence that could appear according to field theory arguments would be at $L=7$ loops. Such a counterterm was explicitly constructed in the linearised theory many years ago \cite{Howe:1980th}, but this is not invariant under the non-linear $E_7$ symmetry. However, there is a seven-loop $E_7$ invariant given by the volume of the on-shell $N=8$ superspace. Although it is known that the volume of superspace can vanish in some lower $N$ examples, there does not seem to be any obvious reason why this should be the case in $N=8$.

\section*{Acknowledgements}

We are grateful to the authors of refs \cite{BernYM,Bern:2009kd}, as well as to Pierre Vanhove, for stimulating discussions. We would also like to thank Marc Henneaux for helpful comments on cohomology.

\end{document}